\documentclass[twocolumn]{revtex4}

\usepackage{ragged2e}
\usepackage{amsmath}
\usepackage{graphicx}
\usepackage{bm}
\usepackage[usenames,dvipsnames]{color}

\definecolor{darkblue}{RGB}{0,0,196}

\begin{document}

\title{Extracting Kinetic Freeze-out Properties in High Energy
Collisions Using a Multi-source Thermal Model} \vspace{0.5cm}

\author{Jia-Yu~Chen$^{1,}${\footnote{E-mail: 202012602001@email.sxu.edu.cn}},
Mai-Ying~Duan$^{1,}${\footnote{E-mail: duanmaiying@sxu.edu.cn}},
Fu-Hu~Liu$^{1,}$\footnote{Correspondence: fuhuliu@163.com;
fuhuliu@sxu.edu.cn},
Khusniddin~K.~Olimov$^{2,3,}$\footnote{Correspondence:
khkolimov@gmail.com; kh.olimov@uzsci.net}}

\affiliation{$^1$State Key Laboratory of Quantum Optics and
Quantum Optics Devices, Institute of Theoretical Physics, Shanxi
University, Taiyuan 030006, China
\\
$^2$Laboratory of High Energy Physics, Physical-Technical
Institute of Uzbekistan Academy of Sciences, Chingiz Aytmatov Str.
2b, Tashkent 100084, Uzbekistan
\\
$^3$Department of Natural Sciences, National University of Science
and Technology MISIS (NUST MISIS), Almalyk Branch, Almalyk 110105,
Uzbekistan}

\begin{abstract}

\vspace{0.5cm}

\noindent {\bf Abstract:} We study the transverse momentum ($p_T$)
spectra of neutral pions and identified charged hadrons produced
in proton--proton ($pp$), deuteron--gold ($d$--Au), and gold--gold
(Au--Au) collisions at the center of mass energy
$\sqrt{s_{NN}}=200$ GeV. The study is made in the framework of a
multi-source thermal model used in the partonic level. It is
assumed that the contribution to the $p_T$-value of any hadron
comes from two or three partons with an isotropic distribution of
the azimuthal angle. The contribution of each parton to the
$p_T$-value of a given hadron is assumed to obey any one of the
standard (Maxwell-Boltzmann, Fermi-Dirac, and Bose-Einstein)
distributions with the kinetic freeze-out temperature and average
transverse flow velocity. The $p_T$-spectra of the final-state
hadrons can be fitted by the superposition of two or three
components. The results obtained from our Monte Carlo method are
used to fit the experimental results of the PHENIX and STAR
Collaborations. The results of present work serve as a suitable
reference baseline for other experiments and simulation studies.
\\
\\
\end{abstract}

\maketitle

\parindent=15pt

\section{Introduction}

The transverse momentum ($p_T$) distributions of the identified
hadrons in the final-state of high energy collisions reflect the
excitation degree of the particle emission source as well as the
speed of collective motion of the particles. The excitation degree
of particle emission source reflects the speed of thermal motion
of the particles. Different distribution functions can be
potential candidates for the description of the $p_T$-spectra
measured in experiments. The distributions include, but are not
necessarily limited to, the standard (Maxwell-Boltzmann,
Fermi-Dirac, and Bose-Einstein) distributions obtained from the
Boltzmann-Gibbs statistics, the Tsallis distribution obtained from
the Tsallis statistics~\cite{3e,3ee,16g,16h,16i,16j}, the Tsallis
form of the standard distribution (or the Tsallis-standard
distribution), the $q$-dual distribution obtained from the
$q$-dual statistics~\cite{52}, the $q$-dual form of the standard
distribution (or the $q$-dual-standard distribution), the Erlang
distribution obtained from the multi-source thermal
model~\cite{12,13,14,15,16}, the Hagedorn function~\cite{16a}
(inverse power law~\cite{16b,16c,16d,16e,16f}) obtained from the
quantum chromodynamics (QCD) calculus etc., and their
superposition.

The flow effect may cause a red shift of the $p_T$-spectra. The
influence of flow effect is not excluded in the above mentioned
distributions~\cite{3e,52,12,16a,16b}. The temperature parameters
used in these distributions are the effective temperature ($T$) of
particle emission source which is larger than the real temperature
one wants to extract from the $p_T$-spectra. Generally, $T$
contains the contributions of thermal motion and collective motion
(flow effect) which are described by the kinetic freeze-out
temperature ($T_0$) and average transverse flow velocity
($\langle\beta_t\rangle$), respectively. The contributions of both
the thermal and collective motions to $p_T$ are not exactly
separable, though the magnitudes of the two effects can be
approximately calculated by some model methods.

One has at least four methods to separate the two types of
motions. {\it The method 1)}: one may use the blast-wave model
with the Boltzmann-Gibbs statistics or Tsallis statistics, in
which a determined velocity profile is assumed, and $T_0$ and
$\langle\beta_t\rangle$ can be obtained
simultaneously~\cite{3a,3b,3c,3d}. The temperature given by the
blast-wave model with the Boltzmann-Gibbs statistics is larger
than that with the Tsallis statistics. {\it The method 2)}: one
may use the Lorentz-like transformation in a $p_T$ distribution,
in which the collective motion is treated as the motion of the
reference system, and $T_0$ and $\langle\beta_t\rangle$ can be
also obtained simultaneously~\cite{3f,3g,3h,3i,3i1,3i2}. {\it The
method 3)}: one may use the intercept--slope method in which $T_0$
is the intercept in the linear relation of $T$ versus the rest
mass ($m_0$) of particle, and $\langle\beta_t\rangle$ is the slope
in the linear relation of the average $p_T$ ($\langle p_T\rangle$)
versus the average energy [i.e., the average mass ($\overline{m}$)
of moving particles in the source rest
frame]~\cite{3j,3k,3l,3m,3n,3o,3p,3q,3q1}. Here, $m_0$ refers to a
given kind of particle in all components in the data sets. In
addition, in the linear relation of $T$ versus $m_0$,
$\langle\beta_t\rangle$ is related to, but not equal to, the
slope~\cite{3r,3s}. {\it The method 4)}: one may propose the
contribution fractions of thermal and collective motions to
$\langle p_T\rangle$ to be determinate values which are from some
models. For example, empirically, $T_0=\langle p_T\rangle/3.07$ in
the hydrodynamic simulations~\cite{3t}, and naturally one obtains
$\langle\beta_t\rangle=(1-1/3.07)\langle
p_T\rangle/\overline{m}=(2.07/3.07)\langle
p_T\rangle/\overline{m}$.

The above four methods were used in our previous
work~\cite{3i1,3i2,3n,3o,3p,3q,3q1}, though only few distributions
were performed. The results from different methods are
inconsistent in some cases. These inconsistent results appear not
only for the magnitudes but also for the tendencies of $T_0$ and
$\langle\beta_t\rangle$ with increasing the collision energy and
centrality. A robust method should be used to obtain a more
reasonable result. The result of the blast-wave model is model
dependent. Although the third and fourth methods seem to be model
independent, they are hard to connect with the basic physics
processes. Considering the fact that the standard distribution is
the most basic one, which is from the relativistic ideal gas model
in the thermodynamics, one prefers to use it with the Lorentz-like
transformation to extract $T_0$ and $\langle\beta_t\rangle$. Here,
the particles (or partons) emitted from the hot and dense system
are assumed to obey the law of the relativistic ideal gas model,
though the matter contained within the intermediate fireball is
known to behave like a strongly interacting fluid of partons. To
our best knowledge, this direct extraction method, using the
standard distribution, is rarely reported in the literature.

In the framework of multi-source thermal model at the quark or
gluon level~\cite{12,13,14,15,16}, one may use the standard
distribution with $T_0$ and $\langle\beta_t\rangle$ to describe
the behavior of partons. For the production of a given particle of
any type, its contributors contain mainly two or three partons
with isotropic azimuthal angles. Of course, one does not expect
that the single-component non-analytical description based on the
superposition of two or three standard distributions with $T_0$
and $\langle\beta_t\rangle$ by the Monte Carlo method is enough to
fit the experimental data. Considering different violent degrees
of binary nucleon-nucleon process in high energy collisions, two
or three or even more sets of parameters are possible, which
results in a non-single-component distribution. Going from the
binary process with the lowest intensity to the most violent one,
the fractions of the corresponding components are getting smaller
and smaller. This means that the fraction of the first component
with smallest $T_0$ and $\langle\beta_t\rangle$ is the largest.

In the case of using the two-component non-analytical description,
the spectrum in high-$p_T$ region is regarded as the result of
hard scattering process which implies high-$T_0$ and
$\langle\beta_t\rangle$, and the spectrum in low-$p_T$ region is
regarded as the result of soft excitation process which implies
low-$T_0$ and $\langle\beta_t\rangle$. In the case of using the
three-component function, one needs one more component, i.e., the
intermediate-$T_0$ and $\langle\beta_t\rangle$ for the spectrum in
intermediate-$p_T$ region. The multi-component function
corresponds to the multi-region fine structure of
$p_T$-spectra~\cite{12-a,12-b,12-c}, which is a natural result of
the multi-source thermal model~\cite{12,13,14,15,16} if the
standard distributions with different values of parameter $T$ are
used to describe different components.

In this article, the standard distribution with $T_0$ and
$\langle\beta_t\rangle$ will be used to describe the transverse
momentum of partons. The transverse momentum $p_T$ of given
particle is the sum of contributions of two or three partons. The
related calculations are performed in the framework of
multi-source thermal model~\cite{12,13,14,15,16}, where the two-
or three-component description is available. The calculated
results are fitted to the experimental data measured in high
energy collisions by the PHENIX~\cite{55} and STAR~\cite{55m}
Collaborations.

The remainder of this article is structured as follows. The
picture and formalism of the multi-source thermal model are
described in Section 2. Results and discussion are given in
Section 3. In Section 4, we give our summary and conclusions.

\section{Picture of multi-source and formalism of multi-component}

According to the multi-source thermal model~\cite{12,13,14,15,16},
one may assume that there are lots of energy sources to form in
high energy collisions. These energy sources can be quarks and/or
gluons if one studies the production of particles. For a given
particle of any type, its contributors may be generally two (for
mesons) or three (for baryons) energy sources of contributor
partons~\cite{12,14}. The number of contributor partons is the
same as that of constituent quarks of a given hadron. In most of
the cases, the contributions of two or three partons are suitable
to fit the hadronic spectra. If the two or three partons are not
enough in the analysis, one may include the contributions from the
fourth or more partons, which corresponds to the hadronic state of
multiple quarks~\cite{12,14}. Here, the contributor partons refer
to the constituent quarks of identified hadrons. In the case of
studying the spectra of leptons, one may consider two contributor
partons as the energy sources, in which one is from the projectile
and the other is from the target.

In the relativistic ideal gas model, the invariant particle
momentum ($p$) distribution can be given by~\cite{3e}
\begin{align}
E\frac{d^3N}{d^3p} = \frac{gV}{(2\pi)^3}E
\bigg[\exp\bigg(\frac{E-\mu}{T}\bigg)+S\bigg]^{-1},
\end{align}
where $E=\sqrt{p^2+m_0^2}=m_T\cosh y$ is the energy,
$m_T=\sqrt{p_T^2+m_0^2}$ is the transverse mass,
$y=(1/2)\ln[(E+p_z)/(E-p_z)]$ is the rapidity, $p_z$ is the
longitudinal momentum, $N$ is the particle number, $g$ is the
degeneracy factor, $V$ is the volume, $\mu$ is the chemical
potential, and $S=0$, $1$, and $-1$ correspond to the
Maxwell-Boltzmann, Fermi-Dirac, and Bose-Einstein distributions,
respectively.

The density function of momenta is obtained by
\begin{align}
\frac{dN}{dp}=\frac{2gV}{(2\pi)^2} p^2
\bigg[\exp\bigg(\frac{E-\mu}{T}\bigg)+S\bigg]^{-1}.
\end{align}
The unit-density function of transverse momentum and rapidity is
written as~\cite{3e}
\begin{align}
\frac{d^2N}{dp_Tdy} =\frac{gV}{(2\pi)^2}p_T E
\bigg[\exp\bigg(\frac{E-\mu}{T}\bigg)+S\bigg]^{-1}.
\end{align}
The density function of transverse momentums is
\begin{align}
\frac{dN}{dp_T} =\frac{gV}{(2\pi)^2} p_T
\int_{y_{\min}}^{y_{\max}} E
\bigg[\exp\bigg(\frac{E-\mu}{T}\bigg)+S\bigg]^{-1}dy,
\end{align}
where $y_{\min}$ and $y_{\max}$ denote the minimum and maximum
rapidities respectively.

In the near mid-rapidity region, $E=m_T\cosh y\approx m_T$. In the
case of having no collective flow, the transverse momentum
$p'_{ti}$ of the $i$-th parton contributing to the transverse
momentum $p_T$ of the given particle is assumed to obey the
probability density function of the standard distribution
\begin{align}
f_{p'_{ti}}(p'_{ti}) =\frac{1}{n}\frac{dn}{dp'_{ti}} = C
p'_{ti}m'_{ti}
\bigg[\exp\bigg(\frac{m'_{ti}-\mu_i}{T}\bigg)+S\bigg]^{-1},
\end{align}
where $n$ is the number of partons, $C$ is the normalization
constant, $m'_{ti}=\sqrt{p'^2_{ti}+m^2_{0i}}$ is the transverse
mass of the $i$-th parton, $m_{0i}$ is the constituent mass that
is 0.31 GeV/$c^2$ for up and down quarks~\cite{55a}, $\mu_i$ is
the chemical potential of the $i$-th parton that is nearly 0 at
high energy~\cite{55b,55c,55d,55e}. One can obtain easily
$f_{p'_{ti}}(p'_{ti})dp'_{ti}$ from Eq. (5). In addition, if $1/n$
in Eq. (5) is replaced by $1/N_{EV}$, where $N_{EV}$ denotes the
number of events, one may obtain the mean multiplicity of
particles in an event.

It should be noted that Eq. (5) is not the united probability
density function of transverse momentum and rapidity (or
longitudinal momentum), but only the probability density function
of transverse momentum at the mid-rapidity which is the concerned
major region in experiments. From a practical point of view, Eq.
(5) is an approximate expression and easy to use. In addition, the
constituent mass, but not the current mass, of a given quark is
used in Eq. (5) due to the considered quarks being the
constituents of the collision system and produced hadrons.

One may introduce the average transverse flow velocity
$\langle\beta_t\rangle$ and the Lorentz-like factor
$\langle\gamma_t\rangle=1/\sqrt{1-\langle\beta_t\rangle^2}$~\cite{3f,3g,3h,3i,3i1,3i2}
at the parton level. The quantities $m'_{ti}$ and $p'_{ti}$, as
well as the transverse mass $m_{ti}$ and transverse momentum
$p_{ti}$ containing the flow effect, can be transformed into each
other. One has the Lorentz-like transformation
\begin{align}
m'_{ti}&=\langle\gamma_t\rangle(m_{ti}-p_{ti}\langle\beta_t\rangle),\nonumber\\
p'_{ti}&=\langle\gamma_t\rangle|p_{ti}-m_{ti}\langle\beta_t\rangle|,\nonumber\\
dp'_{ti}&=\frac{\langle\gamma_t\rangle}{m_{ti}}(m_{ti}-p_{ti}\langle\beta_t\rangle)dp_{ti},
\end{align}
where the absolute value $|p_{ti}-m_{ti}\langle\beta_t\rangle|$ is
used due to $p'_{pt}$ being positive and
$p_{ti}-m_{ti}\langle\beta_t\rangle$ being possibly negative in
low-$p_{ti}$ region. The Lorentz-like, but not the Lorentz, factor
or transformation is called. The reason is that
$\langle\beta_t\rangle$ and $\langle\gamma_t\rangle$, but not
$\beta_t$ and $\gamma_t$, are used in the analysis.

After the Lorentz-like transformation, the probability density
function, $f_{p'_{ti}}(p'_{ti})$, of $p'_{ti}$ is transformed into
the probability density function, $f_i(p_{ti})$, of $p_{ti}$. The
relation between the two probability density functions is
\begin{align}
f_{p'_{ti}}(p'_{ti})|dp'_{ti}| = f_i(p_{ti})|dp_{ti}|.
\end{align}
From Eqs. (5)--(7), one has
\begin{align}
f_i(p_{ti})={}& f_{p'_{ti}}(p'_{ti})\frac{|dp'_{ti}|}{|dp_{ti}|}\bigg|_{m'_{ti},p'_{ti},dp'_{ti} \rightarrow\ m_{ti},p_{ti},dp_{ti}} \nonumber\\
={}&C\frac{\langle\gamma_t\rangle^3}{m_{ti}}|p_{ti}-m_{ti}\langle\beta_t\rangle|
(m_{ti}-p_{ti}\langle\beta_t\rangle)^2 \nonumber\\
&\times\bigg[\exp\bigg(\frac{\langle\gamma_t\rangle(m_{ti}-p_{ti}\langle\beta_t\rangle)
-\mu}{T_0}\bigg)+S\bigg]^{-1}
\end{align}
in which $T$ is naturally rewritten as $T_0$ due to the
introduction of $\langle\beta_t\rangle$. As the quantities at the
parton level, $T_0$ and $\langle\beta_t\rangle$ may show different
tendencies with centrality. The reason is that multiple scattering
of secondary particles may happened in the participants and
spectators which are centrality dependent.

Equation (8) is obtained from Eq. (5) due to the conversion of
probability densities of transverse momentums in which
$\langle\beta_t\rangle$ and $\langle\gamma_t\rangle$ are
considered. Because of Eq. (5) being only for the case of
mid-rapidity, Eq. (8) is deservedly for the same case. Meanwhile,
the application of $\langle\beta_t\rangle$ and
$\langle\gamma_t\rangle$ at the parton level, but not $\beta_t$
and $\gamma_t$ of each parton, can avoid using too many
parameters. In fact, the kinetic freeze-out temperature and
transverse flow velocity extracted from the data fitting are
usually the average quantities.

In the Monte Carlo calculations, let $R_i$ and $r_i$ denote random
numbers distributed evenly in $[0,1]$. A concrete $p_{ti}$
satisfies the relation
\begin{align}
\int_{0}^{p_{ti}}f_{i}(p''_{ti})dp''_{ti}<R_i<\int_{0}^{p_{ti}+\delta
p_{ti}}f_{i}(p''_{ti})dp''_{ti},
\end{align}
where $p''_{ti}$ denotes the integral variable to differ from the
integral upper limit $p_{ti}$ and $\delta p_{ti}$ denotes a small
amount shift from $p_{ti}$. The contributor partons are assumed to
move isotropically in the transverse plane. To obtain a discrete
azimuthal angle $\phi_i$ that satisfies the isotropic
distribution, we have
\begin{align}
\phi_i=2\pi r_i.
\end{align}

In the transverse plane of the rectangular coordinate system, the
$x$- and $y$-components of the vector $\bm p_{ti}$ of the parton
transverse momentum are
\begin{align}
p_{xi}=p_{ti}\cos\phi_i, \nonumber\\
p_{yi}=p_{ti}\sin\phi_i,
\end{align}
respectively. If $n$ partons contribute to $p_T$, the $x$- and
$y$-components of the vector $\bm p_T$ of the particle transverse
momentum are
\begin{align}
p_{x}=\sum_{i=1}^n p_{ti}\cos\phi_i, \nonumber\\
p_{y}=\sum_{i=1}^n p_{ti}\sin\phi_i,
\end{align}
respectively. Then, we have,
\begin{align}
p_T=\sqrt{\bigg(\sum_{i=1}^n p_{ti}\cos\phi_i\bigg)^2
+\bigg(\sum_{i=1}^n p_{ti}\sin\phi_i\bigg)^{2}}.
\end{align}

In particular, if $n=2$ meaning that two contributor partons taken
part in the formation of a particle, we have
\begin{align}
p_T={}&\sqrt{\bigg(\sum_{i=1}^2 p_{ti}\cos\phi_i\bigg)^2
+\bigg(\sum_{i=1}^2 p_{ti}\sin\phi_i\bigg)^{2}}\nonumber\\
={}&\sqrt{p_{t1}^{2}+p_{t2}^{2}+2p_{t1}p_{t2}\cos|\phi_{1}-\phi_{2}|}.
\end{align}
If $n=3$ meaning that three contributor partons taken part in the
formation of a particle, we have
\begin{align}
p_T={}&\sqrt{\bigg(\sum_{i=1}^3 p_{ti}\cos\phi_i\bigg)^2
+\bigg(\sum_{i=1}^3 p_{ti}\sin\phi_i\bigg)^{2}}\nonumber\\
={}&\big(p_{t1}^{2}+p_{t2}^{2}+p_{t3}^{2}+2p_{t1}p_{t2}\cos|\phi_{1}-\phi_{2}|\nonumber\\
&+2p_{t1}p_{t3}\cos|\phi_{1}-\phi_{3}|+2p_{t2}p_{t3}\cos|\phi_{2}-\phi_{3}|\big)^{1/2}.
\end{align}

The $p_T$-spectra of particles can be divided into two or three
regions. This means that one needs two- or three-component
function to fit the $p_T$-spectra. If the first component
describes the spectra in low-$p_T$ region, which corresponds to
the contribution of soft excitation process, the last component
describes the spectra in high-$p_T$ region which corresponds to
the contribution of hard scattering process. Naturally, the
intermediate component (in three-component function) describes the
spectra in intermediate-$p_T$ region. Generally, at low energy or
for narrow $p_T$-spectra, one or two components are needed. At
high energy or for wide $p_T$-spectra, three or more components
are needed.

In the Monte Carlo calculations, one may obtain the digitized
probability density function,
$f_j(p_T,T_{0j},\langle\beta_t\rangle_j)=(1/N)dN/dp_T|_j$, for the
contribution of the $j$-th component, where $T_{0j}$ and
$\langle\beta_t\rangle_j$ denote the kinetic freeze-out
temperature and average transverse flow velocity corresponding to
the $j$-th component. Due to the multi-component, one has the
$p_T$ distribution measured in experiments to be
\begin{align}
f(p_T)=\frac{1}{N}\frac{dN}{dp_T}=\sum_{j=1}^{K}k_jf_j(p_T,T_{0j},\langle\beta_t\rangle_j),
\end{align}
where $K$ denote the number of components and $k_j$ denote the
contribution fraction of the $j$-th component. The normalization
$\sum_{j=1}^{K}k_j=1$ is naturally obeyed.

In the above discussions, from the physics point of view, the
origin of multiple sources has two meanings. For a given kind of
particle, the multiple sources originate from multiple mechanisms
of interactions or different excitation degrees of the system,
which results in the multi-component distribution. If the
particles in low-, intermediate-, and high-$p_T$ regions are
produced by three different mechanisms or from three excitation
degrees, the multiple sources becomes three sources. For a given
particle, the multiple sources refer to multiple energy sources
when the particle is formed. Generally, two or three energy
sources are considered in the formation of the given particle.

It is noteworthy that the collective motion, which pertains to a
common velocity or momentum of the particles, does not lead to the
kinetic freeze-out temperature. In fact, the temperature is known
to originate only from random thermal motion and reflects the
degree of intensity of the thermal motion. In the present work, to
obtain the kinetic freeze-out temperature, the transverse flow
velocity is introduced to exclude the influence of collective
motion. This treatment method is easy to use in the fit of
experimental data. If the influence of collective motion is not
excluded, i.e., if the transverse flow velocity is not considered,
one will obtain the effective temperature which is larger than the
kinetic freeze-out temperature.

In the present work, to fit the experimental invariant yield,
$(1/2\pi p_T) d^2N/dydp_T$, one needs to structure the relation,
$(1/2\pi p_T)N_0f(p_T)=(1/2\pi p_T)\int [d^2N/dydp_T]dy$, where
$N_0$ is the normalization constant that is generally the area
under the data, $dN/dp_T$. In the fit, $N_0$ is determined by the
data itself and has no relation to the model. Although $N_0$ does
not appear as a parameter in the present work to avoid triviality,
one more ``1" is subtracted when counting the number of degree of
freedom (ndof).

\section{Results and discussion}

As an application of the above extraction method based on the
description of $p_T$ spectra, the invariant yields, \((1/2\pi
p_TN_{EV}) d^2N/dydp_T\), of neutral pions (\(\pi^0\)) produced in
mid-pseudorapidity (\(|\eta|<0.35\)) in gold--gold (Au--Au)
collisions with different centrality percentages at the
center-of-mass energy per nucleon pair \(\sqrt{s_{NN}}=200\)~GeV
are presented in Figure 1, where $y$ is the rapidity as defined
above, the pseudorapidity $\eta=-\ln\tan(\theta/2)$, and $\theta$
denotes the emission angle of the particle. The various symbols
represent the experimental data measured by the PHENIX
Collaboration~\cite{55}. The curves are our results fitted by the
three-component function. For clarity, the samples for different
centrality percentages are re-scaled by different factors as
mentioned in the legends in the figure. The values of the
parameters $T_{0j}$, $\langle\beta_t\rangle_j$, $k_j$, $\chi^{2}$,
and ndof are listed in Table 1, where $j=1$, 2, and 3. Here, $k_1$
($=1-k_2-k_3$) is not listed due to the fact that it can be
obtained from the normalization.

One can see that the PHENIX data are fitted satisfactorily by the
three-component function. Although there is no data in the region
of $p_T>12$ GeV/$c$ in peripheral collisions, one may show an
extension of the fitted curves based on the parameters extracted
from the data in the region $p_T<12$ GeV/$c$. The extension of the
fitted curves in peripheral collisions can be compared with those
in central and semi-central collisions. The wavy structure in each
case is caused by the low statistics in high-$p_T$ region.

Based on Table 1, the dependences of parameters on centrality
percentage $C$ are also shown in Figure 2. From the upper to lower
panels, the dependences are for $T_{0j}$,
$\langle\beta_t\rangle_j$, and $k_j$, respectively. It is observed
that with the decrease in centrality (or with the increase in
centrality percentage) from central to peripheral collisions, the
parameters studied here do not have significant change and no
obvious fluctuations are observed. If there are any fluctuations
they are insignificant. One can say that these parameters are
centrality independent. The reason is that the $p_T$-spectra in
Figure 1 are very similar, if not equal, in the shape in different
centrality intervals in Au--Au collisions. This implies that the
kinetic freeze-out parameters extracted from the spectra of
$\pi^0$ are nearly independent of the centrality.

It should be noted that most of the parameters in Table 1 have the
same uncertainties across the centrality, though one has seen some
differences in the fourth decimal place. Only three decimal places
are kept in the Table. Because of parameter uncertainties being
less than the symbol size, they are not visible in Figure 2.

\begin{figure*}[htb!]
\begin{center}
\includegraphics[width=10.0cm]{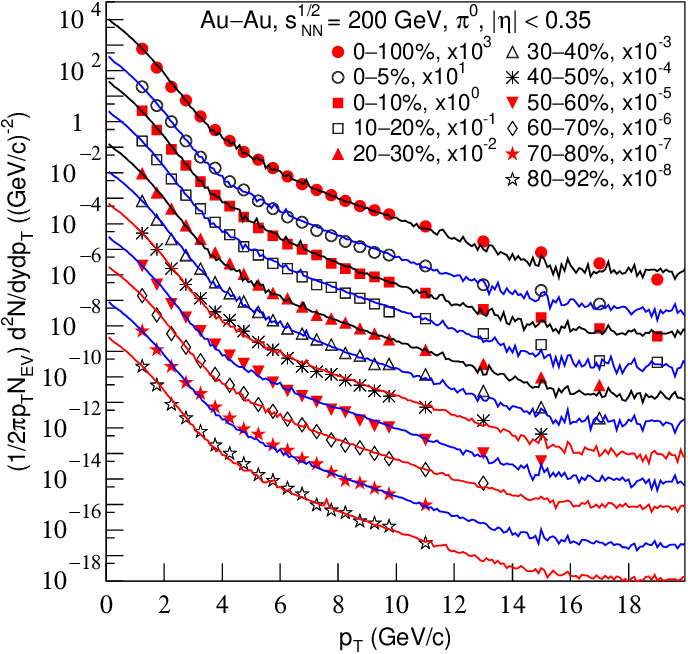}
\end{center}

\justifying\noindent {Figure 1. The invariant yields, \((1/2\pi
p_TN_{EV}) d^2N/dydp_T\), of \(\pi^0\) produced in \(|\eta|<0.35\)
in Au--Au collisions with different centrality percentages at
\(\sqrt{s_{NN}}=200\) GeV. The symbols represent the experimental
data measured by the PHENIX Collaboration~\cite{55}. The curves
are our results fitted by the three-component function in which
each component is regarded as the sum of contributions of two
contributor partons. The contribution of each parton to $p_T$ of
given particle is assumed to obey the standard distribution with
the isotropic azimuthal angle.}
\end{figure*}

\begin{table*}
\vspace{0.5cm} \justifying\noindent {\small Table 1. Values of
$T_{0j}$, $\langle\beta_t\rangle_j$, $k_j$, $\chi^{2}$, and ndof
corresponding to the solid curves in Figure 1 for various
centrality percentages $C$, where $i=1$, 2, and 3. In particular,
$k_1=1-k_2-k_3$ that is not listed.} \vspace{-0.2cm}
{\footnotesize

\begin{center}
\newcommand{\tabincell}[2]
{\begin{tabular}{@{}#1@{}}#2\end{tabular}}
\begin{tabular} {cccccccccc}\\ \hline\hline
$C$ (\%) & $T_{01}$ (GeV) & $T_{02}$ (GeV) & $T_{03}$ (GeV) & $\langle\beta_t\rangle_1$ ($c$) & $\langle\beta_t\rangle_2$ ($c$) & $\langle\beta_t\rangle_3$ ($c$) & $k_2$ (\%) & $k_3$ (\%) & $\chi^2$/ndof \\
\hline
0--100 & $0.178\pm0.002$ & $0.365\pm0.004$ & $0.433\pm0.004$ & $0.239\pm0.002$ & $0.403\pm0.004$ & $0.605\pm0.006$ & $0.160\pm0.002$ & $0.014\pm0.001$ & 24/16\\
0--5   & $0.175\pm0.002$ & $0.366\pm0.004$ & $0.434\pm0.004$ & $0.239\pm0.002$ & $0.403\pm0.004$ & $0.605\pm0.006$ & $0.156\pm0.002$ & $0.015\pm0.001$ & 20/15\\
0--10  & $0.176\pm0.002$ & $0.365\pm0.004$ & $0.432\pm0.004$ & $0.238\pm0.002$ & $0.401\pm0.004$ & $0.610\pm0.006$ & $0.145\pm0.001$ & $0.010\pm0.001$ & 33/16\\
10--20 & $0.175\pm0.002$ & $0.366\pm0.004$ & $0.434\pm0.004$ & $0.239\pm0.002$ & $0.403\pm0.004$ & $0.600\pm0.006$ & $0.145\pm0.001$ & $0.018\pm0.001$ & 35/16\\
20--30 & $0.177\pm0.002$ & $0.363\pm0.004$ & $0.436\pm0.004$ & $0.239\pm0.002$ & $0.405\pm0.004$ & $0.608\pm0.006$ & $0.145\pm0.001$ & $0.018\pm0.001$ & 25/15\\
30--40 & $0.178\pm0.002$ & $0.366\pm0.004$ & $0.433\pm0.004$ & $0.239\pm0.002$ & $0.402\pm0.004$ & $0.602\pm0.006$ & $0.145\pm0.001$ & $0.018\pm0.001$ & 21/15\\
40--50 & $0.178\pm0.002$ & $0.367\pm0.004$ & $0.436\pm0.004$ & $0.239\pm0.002$ & $0.403\pm0.004$ & $0.602\pm0.006$ & $0.152\pm0.002$ & $0.031\pm0.001$ & 17/14\\
50--60 & $0.180\pm0.002$ & $0.366\pm0.004$ & $0.434\pm0.004$ & $0.239\pm0.002$ & $0.402\pm0.004$ & $0.604\pm0.006$ & $0.152\pm0.002$ & $0.034\pm0.001$ & 23/14\\
60--70 & $0.180\pm0.002$ & $0.366\pm0.004$ & $0.431\pm0.004$ & $0.239\pm0.002$ & $0.405\pm0.004$ & $0.605\pm0.006$ & $0.152\pm0.002$ & $0.031\pm0.001$ & 14/13\\
70--80 & $0.181\pm0.002$ & $0.365\pm0.004$ & $0.431\pm0.004$ & $0.240\pm0.002$ & $0.403\pm0.004$ & $0.602\pm0.006$ & $0.165\pm0.002$ & $0.028\pm0.001$ & 15/12\\
80--92 & $0.180\pm0.002$ & $0.367\pm0.004$ & $0.430\pm0.004$ & $0.238\pm0.002$ & $0.403\pm0.004$ & $0.602\pm0.006$ & $0.165\pm0.001$ & $0.028\pm0.002$ & 17/12\\
\hline
\end{tabular}%
\end{center}}
\end{table*}

\begin{table*} \vspace{0.cm} \justifying\noindent {\small Table 2.
Values of correlation coefficients $r_{xy}$ between two
parameters.} \vspace{-0.2cm} {\footnotesize

\begin{center}
\newcommand{\tabincell}[2]
{\begin{tabular}{@{}#1@{}}#2\end{tabular}}
\begin{tabular} {cccccccccc}\\ \hline\hline
 & $T_{01}$ & $T_{02}$ & $T_{03}$ & $\langle\beta_t\rangle_1$ & $\langle\beta_t\rangle_2$ & $\langle\beta_t\rangle_3$ & $k_1$ & $k_2$  &
 $k_3$ \\
\hline
$T_{01}$                  & $1.000$  & $0.174$  & $-0.507$ & $0.266$  & $0.164$  & $-0.242$ & $-0.822$ & $0.611$  & $0.814$ \\
$T_{02}$                  & $0.174$  & $1.000$  & $-0.263$ & $-0.222$ & $-0.296$ & $-0.614$ & $-0.445$ & $0.350$  & $0.423$ \\
$T_{03}$                  & $-0.507$ & $-0.263$ & $1.000$  & $0.202$  & $0.131$  & $0.086$  & $0.397$  & $-0.562$ & $-0.137$ \\
$\langle\beta_t\rangle_1$ & $0.266$  & $-0.222$ & $0.202$  & $1.000$  & $0.320$  & $-0.376$ & $-0.285$ & $0.180$  & $0.312$ \\
$\langle\beta_t\rangle_2$ & $0.164$  & $-0.296$ & $0.131$  & $0.320$  & $1.000$  & $0.003$  & $-0.207$ & $0.080$  & $0.280$ \\
$\langle\beta_t\rangle_3$ & $-0.242$ & $-0.614$ & $0.086$  & $-0.376$ & $0.003$  & $1.000$  & $0.465$  & $-0.349$ & $-0.455$ \\
$k_1$                     & $-0.822$ & $-0.445$ & $0.397$  & $-0.285$ & $-0.207$ & $0.465$  & $1.000$  & $-0.861$ & $-0.875$ \\
$k_2$                     & $0.611$  & $0.350$  & $-0.562$ & $0.180$  & $0.080$  & $-0.349$ & $-0.861$ & $1.000$  & $0.509$ \\
$k_3$                     & $0.814$  & $0.423$  & $-0.137$ & $0.312$  & $0.280$  & $-0.455$ & $-0.875$ & $0.509$  & $1.000$ \\
\hline
\end{tabular}%
\end{center}}
\end{table*}

\begin{figure*}[htb!]
\begin{center}
\includegraphics[width=10.0cm]{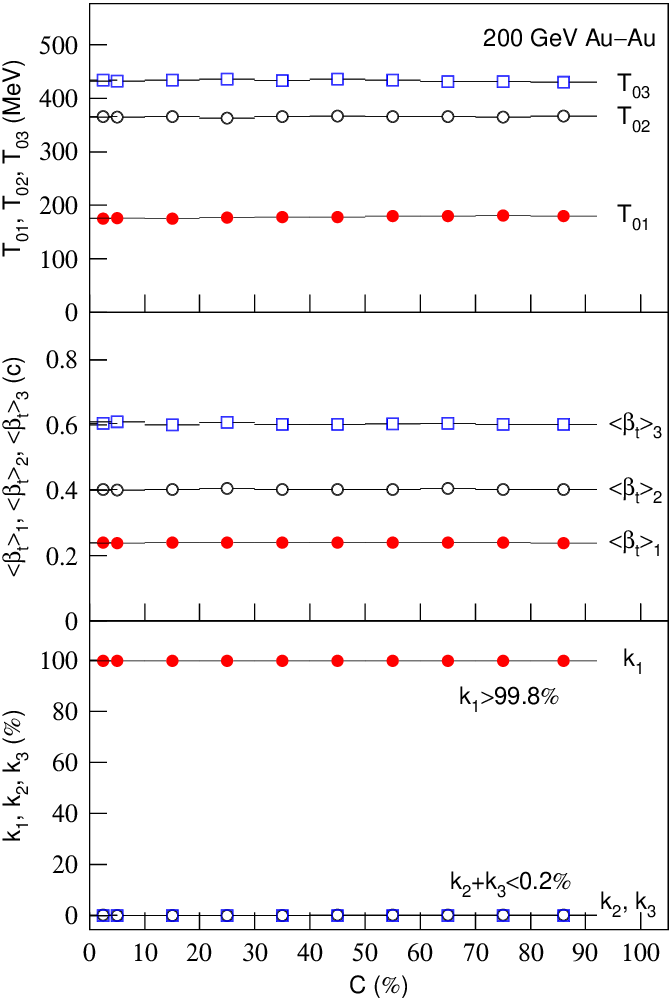}
\end{center}

\justifying\noindent {Figure 2. Dependences of parameters on
centrality percentage $C$ for $\pi^0$ production in Au--Au
collisions at \(\sqrt{s_{NN}}=200\) GeV. From the upper to lower
panels, the dependences are for $T_{0j}$,
$\langle\beta_t\rangle_j$, and $k_j$, respectively. The
contribution ratio of the first component is larger than 99.8\%,
which means that $T_0\approx T_{01}$ and $\langle\beta_t\rangle_1
\approx \langle\beta_t\rangle_{01}$.}
\end{figure*}

\begin{figure*}[htb!]
\begin{center}
\includegraphics[width=10.0cm]{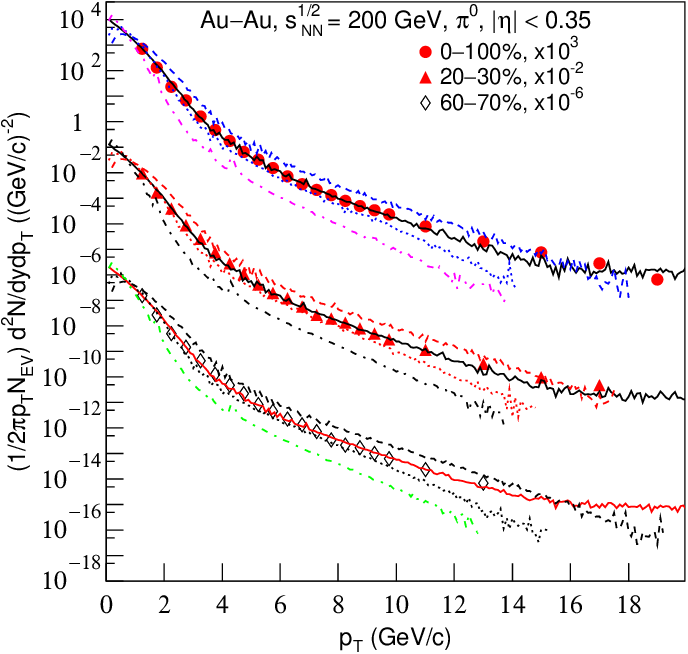}
\end{center}

\justifying\noindent {Figure 3. Comparison of the results of the
isotropic azimuthal angle (the solid curves), the parallel
contributions ($\phi_1=\phi_2$, the dashed curves), the vertical
contributions ($|\phi_1-\phi_2|=\pi/2$, the dotted curves), and
the special case of $\phi_1-\phi_2=3\pi/4$ (the dot-dashed curves)
for three centrality percentages. The tail parts with several
sharp drops in the curves have been cut to avoid confusion. The
symbols and the solid curves are the same as Figure 1.}
\end{figure*}

\begin{figure*}[htb!] \vskip.5cm
\begin{center}
\includegraphics[width=10.0cm]{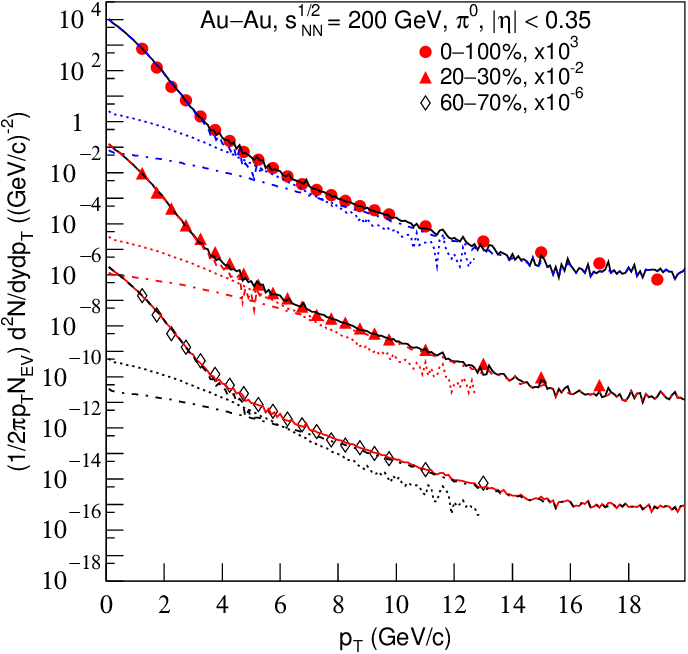}
\end{center}

\justifying\noindent {Figure 4. Comparison of the results from the
three components for three centrality percentages, where the
dashed, dotted, and dot-dashed curves reflect the contributions of
the first, second, and third components, respectively. The tail
parts with several sharp drops in the curves have been cut to
avoid confusion. The symbols and the solid curves are the same as
Figure 1.}
\end{figure*}

\begin{table*}
\vspace{0.5cm} \justifying\noindent {\small Table 3. Values of
$T_{01}$, $\langle\beta_t\rangle_1$, and ndof corresponding to the
curves in Figures 5--7 for Au--Au (upper panel) and $d$--Au
collisions with various centrality percentages $C$ (middle panel),
as well as for $pp$ collisions (lower panel).} \vspace{-0.2cm}
{\footnotesize

\begin{center}
\newcommand{\tabincell}[2]
{\begin{tabular}{@{}#1@{}}#2\end{tabular}}
\begin{tabular} {cc|ccc|ccc|ccc}\\ \hline\hline
 & & $\pi^{\mp}$ & & & $K^{\mp}$ & & & $\bar p(p)$ & & \\
Figure & $C$ (\%) & $T_{01}$ (GeV) & $\langle\beta_t\rangle_1$ ($c$) & $\chi^2$/ndof & $T_{01}$ (GeV) & $\langle\beta_t\rangle_1$ ($c$) & $\chi^2$/ndof & $T_{01}$ (GeV) & $\langle\beta_t\rangle_1$ ($c$) & $\chi^2$/ndof \\
\hline
Figure 5 & 0--5   & $0.132\pm0.012$ & $0.128\pm0.012$ & 14/8 & $0.202\pm0.019$ & $0.245\pm0.021$ & 2/8 & $0.209\pm0.021$ & $0.423\pm0.039$ & 0.1/14 \\
         & 5--10  & $0.132\pm0.012$ & $0.128\pm0.012$ & 11/8 & $0.197\pm0.019$ & $0.241\pm0.021$ & 2/8 & $0.209\pm0.021$ & $0.423\pm0.039$ & 0.2/14 \\
         & 10--20 & $0.132\pm0.012$ & $0.128\pm0.012$ &  8/8 & $0.190\pm0.017$ & $0.233\pm0.020$ & 2/8 & $0.202\pm0.020$ & $0.412\pm0.037$ & 0.2/14 \\
         & 20--30 & $0.131\pm0.012$ & $0.127\pm0.011$ &  7/9 & $0.188\pm0.017$ & $0.231\pm0.020$ & 1/8 & $0.197\pm0.020$ & $0.362\pm0.034$ & 0.1/14 \\
         & 30--40 & $0.130\pm0.012$ & $0.127\pm0.011$ &  5/9 & $0.170\pm0.015$ & $0.210\pm0.018$ & 1/8 & $0.193\pm0.019$ & $0.345\pm0.031$ & 0.2/14 \\
         & 40--50 & $0.127\pm0.011$ & $0.125\pm0.011$ &  3/9 & $0.169\pm0.015$ & $0.185\pm0.016$ & 2/8 & $0.186\pm0.017$ & $0.253\pm0.022$ & 0.1/14 \\
         & 50--60 & $0.124\pm0.010$ & $0.121\pm0.010$ &  2/9 & $0.169\pm0.015$ & $0.185\pm0.016$ & 1/8 & $0.184\pm0.017$ & $0.251\pm0.021$ & 0.1/14 \\
         & 60--70 & $0.123\pm0.009$ & $0.120\pm0.010$ &  1/9 & $0.165\pm0.014$ & $0.145\pm0.013$ & 1/8 & $0.179\pm0.016$ & $0.181\pm0.016$ & 0.1/14 \\
         & 70--80 & $0.120\pm0.008$ & $0.118\pm0.009$ &  1/9 & $0.160\pm0.013$ & $0.118\pm0.008$ & 1/8 & $0.170\pm0.015$ & $0.165\pm0.014$ & 0.1/14 \\
\hline
Figure 6 & 0--100 & $0.119\pm0.008$ & $0.120\pm0.010$ &  4/7 & $0.149\pm0.012$ & $0.180\pm0.016$ & 0.1/8 & $0.167\pm0.014$ & $0.162\pm0.013$ & 0.01/7 \\
         & 0--20  & $0.123\pm0.010$ & $0.122\pm0.011$ &  6/7 & $0.152\pm0.013$ & $0.183\pm0.017$ & 0.2/8 & $0.172\pm0.015$ & $0.167\pm0.014$ & 0.01/7 \\
         & 20--40 & $0.119\pm0.008$ & $0.120\pm0.010$ &  5/7 & $0.149\pm0.012$ & $0.180\pm0.016$ & 0.2/8 & $0.168\pm0.014$ & $0.163\pm0.013$ & 0.01/7 \\
         & 40--100& $0.117\pm0.007$ & $0.118\pm0.009$ &  2/7 & $0.146\pm0.011$ & $0.177\pm0.014$ & 0.1/8 & $0.161\pm0.012$ & $0.155\pm0.011$ & 0.01/7 \\
\hline
Figure 7 & --    & $0.116\pm0.007$ & $0.118\pm0.008$ &  1/9 & $0.155\pm0.012$ & $0.118\pm0.008$ & 0.02/8 & $0.151\pm0.014$ & $0.118\pm0.008$ & 1/13 \\
\hline
\end{tabular}%
\end{center}}
\end{table*}

Table 2 shows the values of correlation coefficients ($r_{xy}$)
between two parameters, where the derived parameter $k_1$ is also
included. It is observed that the correlations are not present for
the case of having no $k_j$ ($|r_{xy}|<0.8$). For example, there
is no significant correlation between $T_{0j}$ and
$\langle\beta_t\rangle_{j'}$, $T_{0j}$ and $T_{0j'}$, as well as
$\langle\beta_t\rangle_{j}$ and $\langle\beta_t\rangle_{j'}$,
where both $j$ and $j'$ are the sequence numbers of the component.
Some correlations involved to $k_j$ are significant
($|r_{xy}|>0.8$). The reason is that the parameters $T_{0j}$ and
$\langle\beta_t\rangle_{j}$ are restrained to the special $p_T$
regions and affect the local shapes of $p_T$-spectra, while the
parameters $k_j$ affect the spectra in whole $p_T$ range due to
the normalization. Concretely, only $T_{01}$ and $k_1$ (or $k_3$
due to $k_3=1-k_1-k_2$), as well as $k_1$ and $k_2$ (or $k_3$) are
highly correlated. This does not give rise to the problem of
multicollinearity.

To see the influence of azimuthal angular difference between the
two contributor partons, Figure 3 displays a comparison of the
results of the isotropic azimuthal angle (the solid curves), the
parallel or identical case ($\phi_1=\phi_2$, the dashed curves),
the vertical case ($|\phi_1-\phi_2|=\pi/2$, the dotted curves),
and the special case of $|\phi_1-\phi_2|=3\pi/4$ (the dot-dashed
curves) for three centrality percentages as an example, where the
tail parts with several sharp drops in the curves have been cut to
avoid confusion. One can see that the shapes of the curves in
low-$p_T$ ($p_T<1$ GeV/$c$) and high-$p_T$ ($p_T>14$ GeV/$c$)
regions are significantly different. In the intermediate-$p_T$
region, the dashed curve is more similar to the solid curve if one
re-normalizes the former. This reflects that in the
intermediate-$p_T$ region one may use the result of
$\phi_1=\phi_2$ to replace approximately that of the isotropic
azimuthal angle. The cases of $|\phi_1-\phi_2|=\pi/2$ and $3\pi/4$
need larger $T_{0j}$ and $\langle\beta_t\rangle_j$ to cater to
that of the isotropic azimuthal angle.

To show the contribution of each component with the isotropic
azimuthal angle, Figure 4 displays the contributions of the first
(the dashed curves), the second (the dotted curves), and the third
(the dot-dashed curves) components together with that of the total
three-component (the solid curves), where the tail parts with
several sharp drops in the curves have been cut to avoid
confusion. Naturally, the first component contributes mainly in
the low-$p_T$ ($p_T<6$ GeV/$c$) region. The second component
contributes mainly in the low- and intermediate-$p_T$ ($p_T<15$
GeV/$c$) regions. The third component contributes in the whole
$p_T$ region. Although the second and third components contribute
in wider $p_T$ region than the first one, the most important
contributor is the first component with $k_1=1-k_2-k_3>0.998$
calculated from Table 1.

Due to very small amounts for $k_{2,3}$ (here $k_2<0.002$ and
$k_3<0.0004$), the second and third components do not affect the
values of $T_0$ ($=k_1T_{01}+k_2T_{02}+k_3T_{03}$) and
$\langle\beta_t\rangle$ ($=k_1\langle\beta_t\rangle_1
+k_2\langle\beta_t\rangle_2 +k_3\langle\beta_t\rangle_3$)
significantly. One has $T_0\approx T_{01}$ and
$\langle\beta_t\rangle \approx \langle\beta_t\rangle_1$. In fact,
including $k_2$ and $k_3$ causes the increase of $T_0$ and
$\langle\beta_t\rangle$ to be less than 0.5\%. One may say that
although there are 8 free parameters in the fit for the data sets
in Figure 1, the main parameters are only $T_{01}$ and
$\langle\beta_t\rangle_1$ which represent reasonably $T_{0}$ and
$\langle\beta_t\rangle$.

In the data analysis, for the purpose of the extraction of kinetic
freeze-out parameters, one does not need too wide $p_T$ spectra.
Generally, the range of $p_T<6$ GeV/$c$ (even $p_T<4$ GeV/$c$) is
enough, though the $p_T$ range is $0\sim20$ GeV/$c$ in some cases
in Figures 1, 3, and 4. If the spectrum in low-$p_T$ region is
contributed by the soft excitation process and that in high-$p_T$
region is contributed by the hard scattering process, the present
work shows that only the contribution of soft process is enough to
extract the kinetic freeze-out parameters. The contribution
fraction of the hard process is smaller than 0.2\%.

The value of $T_0$ is extracted together with
$\langle\beta_t\rangle$ at the parton level. Due to the
introduction of $\langle\beta_t\rangle$, $T_0$ becomes smaller
than the effective temperature $T$ because the flow effect is
excluded. The introduction of $\langle\beta_t\rangle$ also affect
the tendency of $T_0$ with changing the centrality and other
factors. At both the parton and particle levels, the obtained
$T_0$ or $\langle\beta_t\rangle$ may be inconsistent in the
tendency and/or size due to different extraction methods. One
should have a determined and uniform method to extract $T_0$ and
$\langle\beta_t\rangle$. As a tentative work, the present work
focuses on the parameters at the parton level. Meanwhile, the
standard distribution used in the relativistic ideal gas model is
applied for the system. In the aspect for extracting the
parameters at the kinetic freeze-out, this work has an important
significance in methodology.

Many data sets have been analyzed in our previous
work~\cite{3n,3o,3p,3q,3q1} by the thermal-related models, though
only the $p_T$ spectra of $\pi^0$ in Au--Au collisions at
$\sqrt{s_{NN}}=200$ GeV are mainly analyzed in this work as an
example to check the validity of the model in methodology. In our
recent work~\cite{55f}, using the blast-wave model with
fluctuations~\cite{55g,55h,55i}, the kinetic freeze-out parameters
extracted from the $p_T$ spectra of charged pions ($\pi^-+\pi^+$)
and kaons ($K^-+K^+$) produced in lead--lead (Pb--Pb) collisions
at $\sqrt{s_{NN}}=2.76$ TeV~\cite{55j}, proton--lead ($p$--Pb)
collisions at $\sqrt{s_{NN}}=5.02$ TeV~\cite{55k}, and
xenon--xenon (Xe--Xe) collisions at $\sqrt{s_{NN}}=5.44$
TeV~\cite{55l} show almost the independence of centrality, and the
parameters extracted from the $p_T$ spectra of anti-protons and
protons $(\bar p+p)$ show the dependence of centrality.

The difference between the parameters from the spectra of mesons
and $\bar p+p$ is caused by different production mechanisms.
Generally, mesons are newly produced and some protons already
exist in the projectile and target nuclei before the collisions.
In our opinion, the extracted parameter values for $\bar p+p$
production are reduced by the influence of these pre-existing
protons, which are the leading particles with low excitation in
the collisions. They increase relatively the yield in low-$p_T$
region. The relative increase is more obvious in peripheral
collisions due to lesser multiple scattering.

\begin{figure*}[htb!]
\begin{center}
\includegraphics[width=14.0cm]{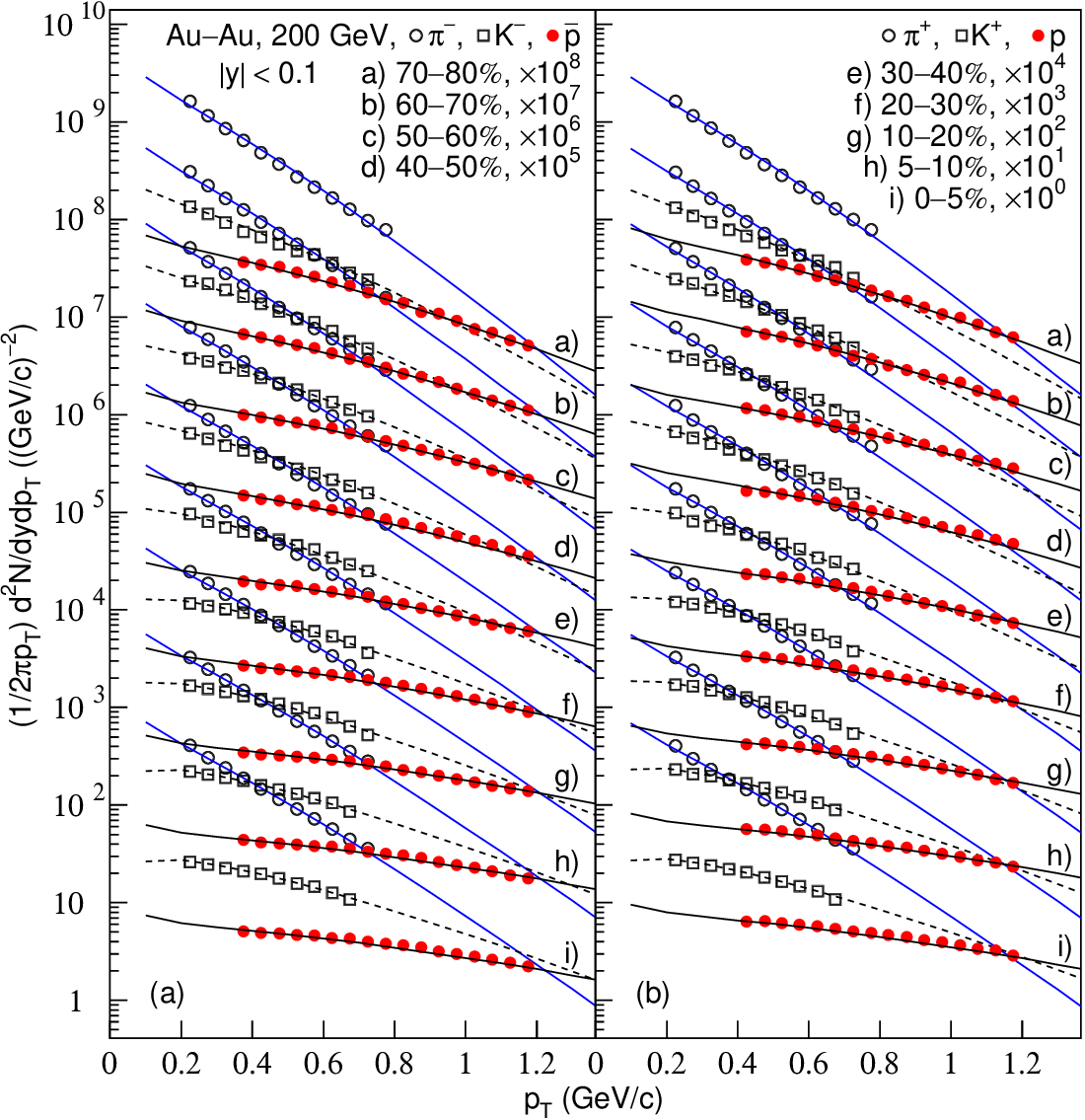}
\end{center}

\justifying\noindent {Figure 5. The invariant yields, \((1/2\pi
p_T) d^2N/dydp_T\), of \(\pi^-\), \(\pi^+\), \(K^-\), \(K^+\),
\(\bar p\), and \(p\) produced in \(y<0.1\) in Au--Au collisions
with different centrality percentages at \(\sqrt{s_{NN}}=200\)
GeV. The symbols represent the experimental data measured by the
STAR Collaboration~\cite{55m}. The curves are our results fitted
by the single-component function (or the first component in the
three-component function), where the sum of contributions of two
(for mesons) or three (for baryons) contributor partons is
considered.}
\end{figure*}

\begin{figure*}[htb!]
\begin{center}
\includegraphics[width=12.0cm]{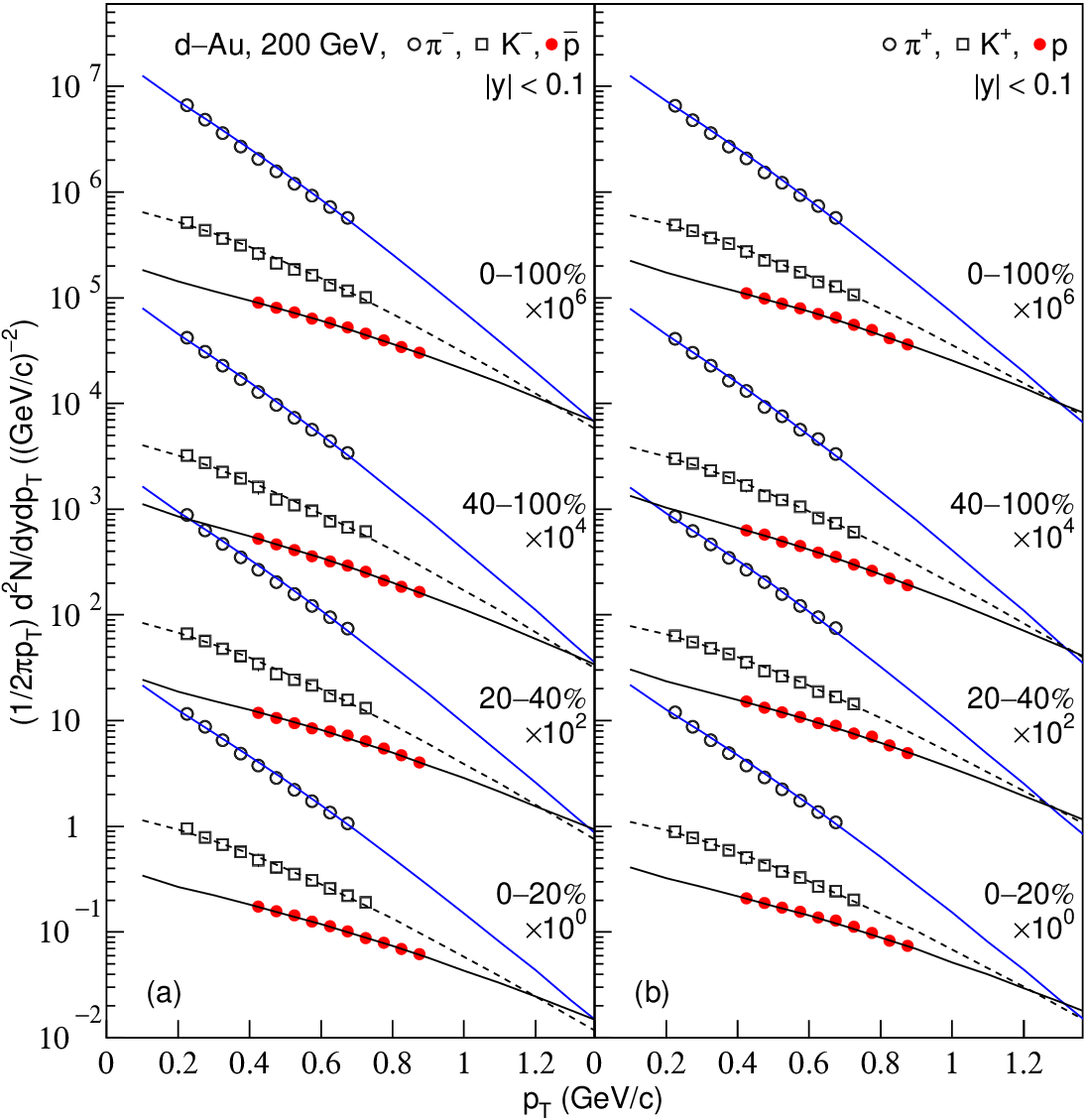}
\end{center}

\justifying\noindent {Figure 6. The invariant yields, \((1/2\pi
p_T) d^2N/dydp_T\), of \(\pi^-\), \(\pi^+\), \(K^-\), \(K^+\),
\(\bar p\), and \(p\) produced in \(y<0.1\) in \(d\)--Au
collisions with different centrality percentages at
\(\sqrt{s_{NN}}=200\) GeV. The symbols represent the experimental
data measured by the STAR Collaboration~\cite{55m}. The curves are
our results fitted by the single-component function.}
\end{figure*}

\begin{figure*}[htb!]
\begin{center}
\includegraphics[width=12.0cm]{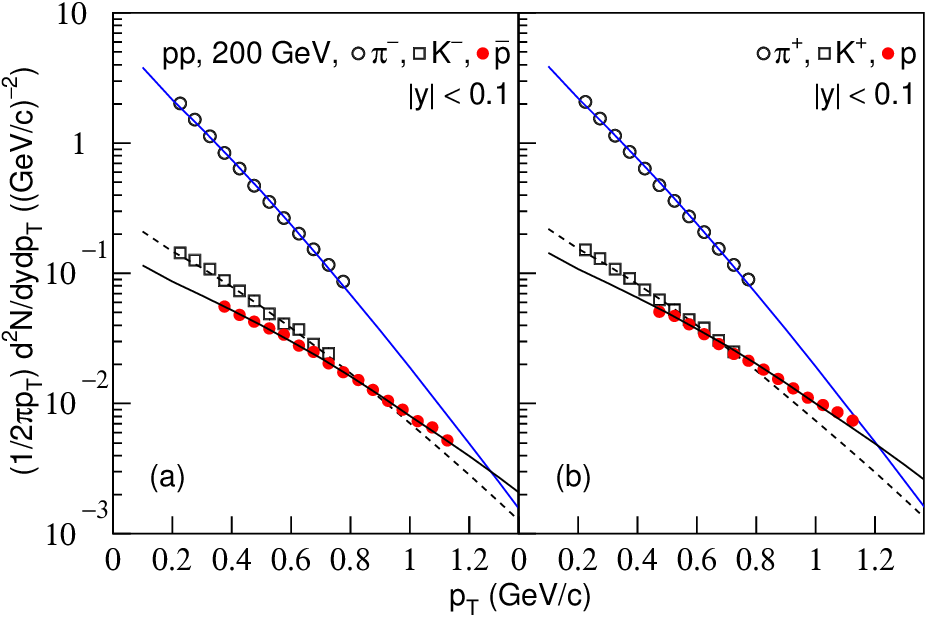}
\end{center}

\justifying\noindent {Figure 7. The invariant yields, \((1/2\pi
p_T) d^2N/dydp_T\), of \(\pi^-\), \(\pi^+\), \(K^-\), \(K^+\),
\(\bar p\), and \(p\) produced in \(y<0.1\) in \(pp\) collisions
at \(\sqrt{s}=200\) GeV. The symbols represent the experimental
data measured by the STAR Collaboration~\cite{55m}. The curves are
our results fitted by the single-component function.}
\end{figure*}

\begin{figure*}[htb!] \vskip.5cm
\begin{center}
\includegraphics[width=10.0cm]{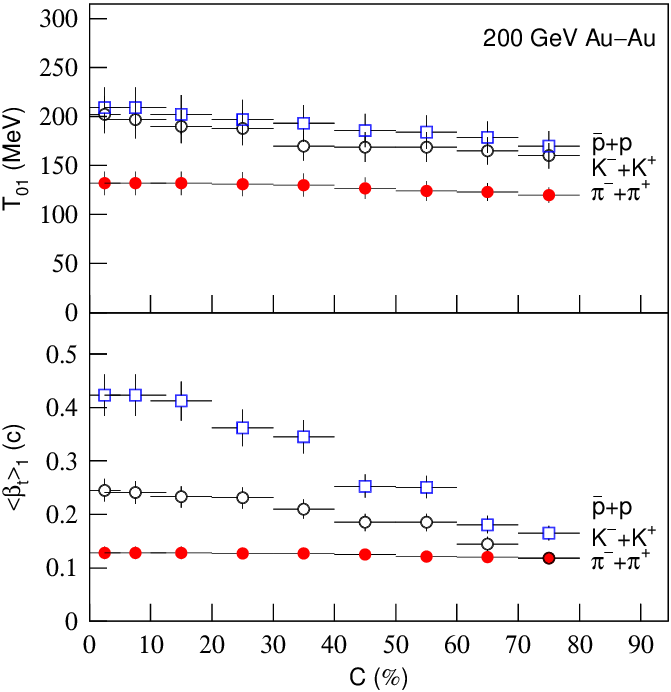}
\end{center}

\justifying\noindent {Figure 8. Dependences of $T_{01}$ (upper
panel) and $\langle\beta_t\rangle_1$ (lower panel) on centrality
percentage $C$ and particle kind. In each panel, the closed
circles, open circles, and open squares represent the results for
$\pi^-+\pi^+$, $K^-+K^+$, and $\bar p+p$, respectively.}
\end{figure*}

It is expected that for a narrow spectrum in low-$p_T$ region, a
single- or two-component function is suitable; for a not too wide
$p_T$ spectrum, the three-component function is suitable; and for
a wider $p_T$ spectrum, the fourth component may be included if
the three components are not enough. To check the methodology by
the spectra in a narrow $p_T$ region, for which only one component
is used in the fit, Figures 5(a)--7(a) shows the spectra of
$\pi^-$, $K^-$, and $\bar p$, and Figures 5(b)--7(b) shows the
spectra of $\pi^+$, $K^+$, and $p$, produced in Au--Au,
deuteron--gold ($d$--Au), and proton--proton ($pp$) collisions at
$\sqrt{s_{NN}}=200$ GeV, where $\sqrt{s_{NN}}$ is simplified to
$\sqrt{s}$ for $pp$ collisions. The symbols represent the
experimental data measured by the STAR Collaboration~\cite{55m}.
The curves are our fitted results by a single-component function
in which only $T_{01}$ and $\langle\beta_t\rangle_1$ are free
parameters which are listed in Table 3. For the productions of
$\pi^-+\pi^+$ and $K^-+K^+$, two contributor partons contribute to
$p_T$. For the production of $\bar p+p$, three contributor partons
contribute to $p_T$. One can see that the single-component
function can fit the spectra in the narrow $p_T$ region. Note that
some of the $\chi^2$/ndof values in Table 3 are very small. This
is because the combined errors in $p_T$ spectra are predominantly
determined by large systematic uncertainties, which do not follow
a normal distribution. Whereas the statistical uncertainties,
which have a random nature, are very small and negligible.

The parameter values listed in Table 3 show that $pp$ collisions
are similar to peripheral $d$--Au and Au--Au collisions at the
same $\sqrt{s_{NN}}$. Central, semi-central, and peripheral
$d$--Au collisions are similar to peripheral Au--Au collisions.
Larger values of parameters in central Au--Au collisions imply
that larger amounts of collision energy is deposited in the system
and more multiple scatterings occur in the hot and dense matter.

Table 3 also shows that $T_{01}$ and $\langle\beta_t\rangle_1$
decrease with the decrease in centrality from central to
peripheral events, and decrease with the decrease in particle mass
for the considered three kinds of particles, in Au--Au and $d$--Au
collisions at $\sqrt{s_{NN}}=200$ GeV. Figure 8 shows the
dependence of $T_{01}$ and $\langle\beta_t\rangle_1$ on centrality
percentage in Au--Au collisions. One can see clearly the
dependence of the parameters on centrality and particle kind.
Comparing with that for $\pi^0$, the dependence for charged
massive particles ($K^-+K^+$ and $\bar p+p$) is significant and
the distributions for $\pi^-+\pi^+$ are almost flat, also showing
no dependence on centrality. The reason is that charged massive
particles have more probability to interact with spectator
nucleons, which causes a large red shift of the spectra in
peripheral collisions.

Generally, $T_0$ and $\langle\beta_t\rangle$ reflect the
excitation and expansion degrees of the system respectively. The
larger these parameters are, the higher the excitation and
expansion degrees of the system is. In the case of charged massive
particles, central collisions correspond to higher excitation and
expansion degrees due to larger amounts of collision energy being
deposited. Meanwhile, charged massive particles have relative
large elastic scattering cross-section due to their large sizes.
While, neutral and charged pions have very small elastic
scattering cross-section in the system, which results their
parameters showing almost independent of centrality.

The reason why one may apply the same distribution to the energy
sources with different degrees of excitation and expansion in high
energy collision systems is because of the similarity,
commonality, and universality, especially the universality, in the
collisions~\cite{4,5,6,7,8,9,10,11}. In particular, in high energy
collisions, the underlying reason is the contributor partons
appearing as the energy sources or influence factors. This also
explains the consistency of some quantities in high energy
collision systems with different sizes and centralities. Some
model independent dependencies, if available, are mainly caused by
the effective energies of the contributor partons or energy
sources.

In the above discussion, the classical concepts of temperature and
equilibrium are tentatively used. However, the collision system is
very small. In particular, only two or three contributor partons
are considered in the formation of given particle. It seems that
the mentioned concepts are not applicable. In fact, although few
partons are considered to contribute directly and mainly to
particle's $p_T$, lots of partons exist in the system of high
energy collisions. In addition, the experiments of high energy
collisions involve lots of events. One may use the grand canonical
ensemble to study the particle production. Then, the concepts of
temperature and equilibrium are applicable. At least, one may
regard $T_0$ and $\langle\beta_t\rangle$ as parameters that
describe the average kinetic energies of thermal and collective
motions of partons respectively.

Before summary and conclusions, we would like to emphasize that
although many parameters are used in this article, this is only
for the wide $p_T$ spectra with multi-region structure. Indeed,
mathematically, increasing the number of parameters increases the
probability of a good fit as one has more free parameters to play
around. Technically, one may not completely discard the usage
because there are different particle production mechanisms in
different $p_T$ regions. This work uses the standard distribution
for a given parton source, and the multi-component distribution
for the parton sources with different excitation degrees. In the
case of extracting $T_0$ and $\langle\beta_t\rangle$, only two
parameters $T_{01}$ and $\langle\beta_t\rangle_1$ are enough. The
distribution with $T_{01}$ and $\langle\beta_t\rangle_1$ ($\neq0$)
describes wider spectra than that with only $T_{01}$ (where
$\langle\beta_t\rangle_1=0$). Because this work is based on the
standard distribution which is widely used in statistical analysis
in modern physics, the results are suitable to be the baseline for
comparing with other experiments and simulation studies.

In addition, although it has been established for almost 20 years
that the high-$p_T$ spectra of hadrons in nuclear collisions are
explained by jet quenching, i.e., high-$p_T$ partons lose energy
and then fragment to hadrons. This work shows an alternative and
uniform explanation for the statistical behavior of particle
spectra in various $p_T$ regions. There is no contradiction
between the two explanations. If the explanation of jet quenching
is focused on the production mechanism, the present work is
focused on the statistical law obeyed by the contributor partons
and produced particles. Even if in the high-$p_T$ region, the
standard distribution is applicable to extract the kinetic
freeze-out parameters. Although this results in quite a large
$T_{02}$ and $T_{03}$ ($\sim0.365-0.435$ GeV) in the fitting for
all centralities, the value of $T_0$ weighted by $T_{0i}$ is
small. Here, the weighted factor is $k_i$. $T_0$ is determined by
$T_{01}$ that is extracted from the pion spectra to be
$\sim0.132-0.175$ GeV in central Au--Au collisions, $\sim0.123$
GeV in central $d$--Au collisions, and $~0.116$ GeV in $pp$
collisions at $\sqrt{s_{NN}}=200$ GeV. It is likely that the
unexpected large $T_{02}$ and $T_{03}$ are obtained by the hard
process due to violent head on collisions between partons, though
this probability is very small.

Different methods or functions used in the extractions of
temperature and flow velocity are different ``thermometers" and
``speedometers". Although the tendencies of parameters based on
different methods are almost the same or approximately the same,
there are differences in concrete values. Obviously, before giving
a comparison, these thermometers and speedometers should be
uniformed according to the selected baseline. In our opinion, the
standard distribution is a good candidate to be the baseline. The
fact that $T_{0j}$ and $\langle\beta_t\rangle_j$ for pion
production are almost independent of centrality is an indicator
that there is something more universal than bulk medium flow that
governs the physics in different $p_T$ regions. The underlying
contributors are partons, but not nucleons, in various collisions
at high energy. This also implies the similarity, commonality, and
universality, in particular universality, in high energy
collisions~\cite{4,5,6,7,8,9,10,11}

\section{Summary and conclusions}

In the framework of multi-source thermal model used in the parton
level, the transverse momentum spectra of the final-state neutral
pions and identified charged hadrons produced in
mid-(pseudo)rapidity region in Au--Au and $d$--Au collisions with
various centralities and in $pp$ collisions at $\sqrt{s_{NN}}=200$
GeV have been studied. For a given particle of any type, its
contributors may be two or three partons with isotropic azimuthal
angle distribution. The contribution of each parton to transverse
momentum of the hadron is assumed to obey the standard
distribution with given kinetic freeze-out temperature and average
transverse flow velocity. The transverse momentum spectra of the
final-state hadrons can be fitted by the superposition of two or
three components. The number of components is related to the width
of transverse momentum spectra.

The results calculated by the Monte Carlo method fit
satisfactorily the experimental data measured by the PHENIX and
STAR Collaborations. With the decrease in centrality from central
to peripheral collisions, the kinetic freeze-out temperature and
average transverse flow velocity for each component in pion
production do not change significantly. Due to the very small
contribution fractions of the second and third components, the
main parameters are determined by the first component in the low
transverse momentum region. The result corresponding to the
isotropic azimuthal angles is similar to that of the identical
azimuthal angles. The kinetic freeze-out parameters decrease with
the decrease in centrality for the production of the charged
massive hadrons. The work based on the standard distribution is
suitable to be the baseline in comparing with other experiments
and simulation studies.
\\
\\
{\bf Data Availability}

The data used to support the findings of this study are included
within the article and are cited at relevant places within the
text as references.
\\
\\
{\bf Ethical Approval}

The authors declare that they are in compliance with ethical
standards regarding the content of this paper.
\\
\\
{\bf Disclosure}

The funding agencies have no role in the design of the study; in
the collection, analysis, or interpretation of the data; in the
writing of the manuscript; or in the decision to publish the
results.
\\
\\
{\bf Conflicts of Interest}

The authors declare that there are no conflicts of interest
regarding the publication of this paper.
\\
\\
{\bf Acknowledgments}

This paper applies the multi-source thermal model, which has also
been applied in our recent work~\cite{67,68}. As a result, some
similar statements are inevitably applied in this paper. The work
of Shanxi Group was supported by the National Natural Science
Foundation of China under Grant No. 12147215, the Shanxi
Provincial Natural Science Foundation under Grant No.
202103021224036, and the Fund for Shanxi ``1331 Project" Key
Subjects Construction. The work of K.K.O. was supported by the
Agency of Innovative Development under the Ministry of Higher
Education, Science and Innovations of the Republic of Uzbekistan
within the fundamental project No. F3-20200929146 on analysis of
open data on heavy-ion collisions at RHIC and LHC.
\\
\\

\end{document}